\documentclass{ifacconf}

\usepackage{graphicx}      
\usepackage{xcolor}
\usepackage{amsfonts}
\usepackage{siunitx}
\usepackage[utf8]{inputenc}
\usepackage{mathtools}
\usepackage{natbib}
\usepackage{makecell}
\usepackage{booktabs}

\begin{document}
\begin{frontmatter}

\title{Distributionally Robust Model Predictive Control for Virtual Power Plants} 
\thanks[footnoteinfo]{This work was supported by FME Solar, funded by the Research Council of Norway and industry partners, under project number 350244.}

\author[First]{Nikolas Recke} 
\author[First]{Mathias Hudoba de Badyn} 

\address[First]{Department of Technology Systems,    University of Oslo, Kjeller, Norway (e-mail: \{nikolalr, mathihud\}@uio.no).}

\begin{abstract}
This paper presents a distributionally robust model predictive control (DR-MPC) framework for the optimal Virtual Power Plant (VPP) operation under electricity price uncertainty. A unified VPP model is formulated that captures the interaction between buildings, battery storage, and renewable generation, all influenced by exogenous weather and market signals. The proposed approach integrates data-driven forecasting with quantile-based uncertainty quantification to construct time-varying Wasserstein ambiguity sets that adapt to forecast dispersion and distributional shifts. This yields a tractable DR-MPC formulation that incorporates predictive distribution information directly into real-time decision making. The method is evaluated using real weather and market data from a Nordic case study across two seasonal scenarios. The results show that DR-MPC improves economic performance relative to standard forecast-based MPC when the ambiguity radius is chosen appropriately, with consistent gains of up to 0.8\,\% for small radii across both seasonal scenarios. Larger radii become overly conservative and reduce revenue, underscoring the importance of proper radius selection. These findings demonstrate the practical value of distributionally robust optimization for 
uncertainty-aware VPP operation.
\end{abstract}


\end{frontmatter}

\section{Introduction}
Renewable energy sources (RES) are playing an increasingly important role in global electricity generation, but their inherent variability introduces uncertainty in power generation that complicates grid stability and system operation (\cite{Zuo2023, Marinescu2022}). In energy systems that additionally face uncertain electricity market prices as well as uncertainties in renewable generation and thermal demand, these uncertainties pose major challenges for optimal operational decision-making (\cite{Morales2014}). Forecast errors in these variables can propagate through the control system, leading to suboptimal scheduling, increased costs, or constraint violations, therefore modern control schemes must  explicitly account for uncertainty.

Virtual Power Plants (VPPs) aggregate dispatchable and non-dispatchable sources, residential buildings, and battery storage into a single controllable entity capable of providing grid support and participating in electricity markets (\cite{Marinescu2022}). However, effective VPP operation requires predictive, uncertainty-aware control strategies that handle the stochastic nature of both physical and market environments.

The operation of VPPs under uncertain forecasts constitutes a decision-making under uncertainty problem for which several control paradigms have been proposed. Model Predictive Control (MPC) provides some inherent robustness through its receding-horizon structure, but its performance deteriorates under large, unmodeled disturbances (\cite{Camacho2007}). Robust MPC (RMPC) ensures constraint satisfaction for all admissible disturbances yet is often overly conservative, while Stochastic MPC (SMPC) reduces conservatism but depends on precise probabilistic knowledge rarely available in practice (\cite{McAllister2024a}). Distributionally Robust Optimization (DRO) offers a data-driven alternative that optimizes against all distributions within a data-informed ambiguity set, providing robustness to distributional shift and model misspecification (\cite{MohajerinEsfahani2017}). DR-MPC bridges the gap between SMPC and RMPC by combining robustness, interpretability, and adaptability (\cite{McAllister2024a}).

The main contributions of this paper are as follows:
\begin{itemize}
    \item formulation of a distributionally robust optimal control problem for real-time VPP operation, jointly optimizing economic performance and thermal comfort while explicitly robustifying with respect to electricity price uncertainty,
    \item construction of time-varying Wasserstein ambiguity sets from empirical quantile forecasts of the market price,
    \item integration within a receding-horizon DR-MPC framework using causal probabilistic forecasts from the TiREx forecaster,
    \item validation on a realistic Nordic case study using real weather and market data, demonstrating consistent revenue improvements over forecast-based MPC.
\end{itemize}

\section{Virtual Power Plant Model}
\label{vpp}

\begin{figure}[ht!]
    \centering
    \includegraphics[width=0.85\linewidth]{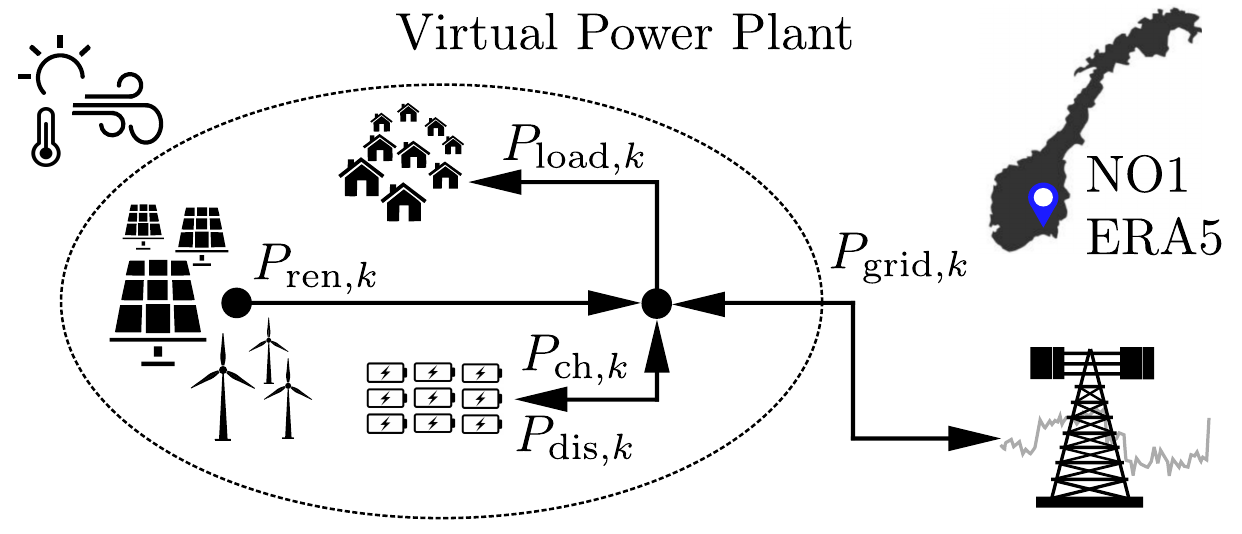}
    \caption{Conceptual illustration of a Virtual Power Plant.}
    \label{fig:vpp_fig}
\end{figure}

The VPP considered in this work aggregates residential buildings, battery storage systems, renewable generation through wind and solar plants, and a grid interface into a single controllable entity, see Figure \ref{fig:vpp_fig}.

At each time step $k$, the VPP must satisfy an instantaneous power balance that links renewable production, building demand, battery operation, and grid exchange. The power balance at each time step is
\begin{align}
    P_{\text{ren},k} - P_{\text{load},k} - P_{\text{bat},k} = P_{\text{grid},k},
    \label{eq:power-balance}
\end{align}

We represent the battery by a net electrical power $P_{\text{bat},k} = P_{\text{ch},k} - P_{\text{dis},k}$,
where $P_{\text{ch},k} \geq 0$ and $P_{\text{dis},k} \geq 0$ denote the charging and discharging powers respectively, and 
positive values of $P_{\text{bat},k}$ correspond to charging. Similarly, the grid exchange is expressed as $P_{\text{grid},k} = P_{\text{exp},k} - P_{\text{imp},k}$, where $P_{\text{exp},k} \geq 0$ and $P_{\text{imp},k} \geq 0$ denote the power exported to and imported from the grid. For tractability, import and export are represented as separate non-negative variables, without explicitly enforcing mutual exclusivity.

\subsection{Energy Storage System}
The dynamics of the battery storage system are modeled by the discrete-time equation 
\begin{align}
S_{k+1} &= S_{k} 
    + \eta^{\text{ch}} P^{\text{ch}}_{k}\tau 
    - \frac{1}{\eta^{\text{dis}}} P^{\text{dis}}_{k}\tau, \label{eq:batt-dyn}
\end{align}
subject to
\begin{align}
S^{\min} \le S_{k} \le S^{\max},\quad
0 \le P^{\text{ch}}_{k} \le \bar P^{\text{ch}},\quad
0 \le P^{\text{dis}}_{k} \le \bar P^{\text{dis}}, \label{eq:batt-limits}
\end{align}
where $S_k$ denotes the state of charge (SoC), $P^{\text{ch}}_k$ and $P^{\text{dis}}_k$ are the charging and discharging powers, $\eta^{\text{ch}}$, $\eta^{\text{dis}}$ the efficiencies, and $\tau$ the time step in hours. 

\subsection{Renewable Energy Generation}
The VPP has access to different renewable assets, such as solar and wind plants.  
The power generated by the photovoltaic (PV) modules at each discrete time step $k$ is modeled as
\begin{align}
P_{\text{solar},k}
    &= \eta_{\text{PV}}\,A_{\text{PV}}\,I_{\text{sol},k}
       \big[1 - \gamma_T\big(T_{\text{cell},k}-T_{\text{ref}}\big)\big], \label{eq:pv}
    \end{align}
where $\eta_{\text{PV}}$ is the nominal efficiency, $A_{\text{PV}}$ the panel area, $I_{\text{sol},k}$ the solar irradiance, and $\gamma_T$ the temperature coefficient.  
The cell temperature is estimated as $T_{\text{cell},k} = T_{\text{air},k} + k_T I_{\text{sol},k}$.

The electrical power output of a wind turbine is given by
\begin{align}
P_{\text{wind},k}
    &= \tfrac{1}{2}\,\rho_{\text{air}}\,A_{\text{rot}}\,\eta_{\text{wind}}\,v_k^3, \label{eq:wind}
\end{align}
where $\rho_{\text{air}}$ is the air density, $A_{\text{rot}} = \pi r_{\text{b}}^2$ the rotor-swept area, $\eta_{\text{wind}}$ the overall efficiency, and $v_k$ the wind speed.  

For the aggregated VPP, the total renewable generation is expressed as
\begin{align}
P_{\text{ren},k}
    &= \sum_{i=1}^{N_S} P_{\text{solar},i,k}
     + \sum_{j=1}^{N_W} P_{\text{wind},j,k}. \label{eq:renew-agg}
\end{align}

\subsection{Energy Demand of Residential Buildings}

The thermal dynamics of a residential building is represented by a three-resistance, two-capacitance (3R2C) model (\cite{Bacher2011}). The model describes the heat exchange between the outer and inner walls and the indoor air temperature. 
The continuous-time dynamics are given by
\begin{align}
\dot T_1 &= -\tfrac{hA + 1/R}{C_1}T_1 + \tfrac{1}{R C_1}T_2
           + \tfrac{hA}{C_1}T_0 + \tfrac{1}{C_1}q_{\text{sol}}A, \nonumber\\
\dot T_2 &= \tfrac{1}{R C_2}T_1 - \tfrac{hA + 1/R}{C_2}T_2 + \tfrac{hA}{C_2}T_z, \nonumber\\
\dot T_z &= \tfrac{hA}{C_z}(T_2 - T_z)
           + \tfrac{1}{C_z}\!\left(Q_{\text{ihg}} + Q_{\text{HVAC}}\right), \label{eq:3r2c}
\end{align}
where $T_1$, $T_2$, and $T_z$ denote the wall and indoor air temperatures, and $T_0$ is the outdoor temperature. The inputs to the 3R2C model are the outdoor air temperature $T_{0}$, the effective solar gains $q_{\text{sol}}A$, the internal heat gains $Q_{\text{ihg}}$, and the HVAC thermal power $Q_{\text{HVAC}}$, which represents the thermal energy exchanged between the HVAC system and the zone.

After discretization with sampling period $\tau$, the system can be written in linear state-space form:
\begin{align}
x_{k+1} = A_d x_k + B_d u_k, \qquad y_k = C x_k, 
\label{eq:building-ss}
\end{align}
where the discrete-time matrices are obtained via zero-order 
hold discretization of the continuous-time 
dynamics~\eqref{eq:3r2c} with sampling period $\tau$, 
specifically $A_d = e^{A_c \tau}$ and 
$B_d = A_c^{-1}(A_d - I)B_c$. 
The state vector is $x_k = [T_1,\,T_2,\,T_z]^\top$ and 
the input vector 
$u_k = [T_0,\,q_{\text{sol}}A,\,Q_{\text{ihg}},\,Q_{\text{HVAC}}]^\top$.
The HVAC electrical power consumption is related to the 
thermal input via $P_{\text{el},k} = Q_{\text{HVAC},k}/\text{COP}$. For each building, the indoor temperature is constrained to remain within a predefined comfort band,
\begin{align}
    T_z^{\min} \le T_{z,k} \le T_z^{\max}, \qquad \forall\, k ,
    \label{eq:building-comfort}
\end{align}
and the HVAC heat flow is limited by its rated capacity,
\begin{align}
    -Q_{\max} \le Q_{\text{HVAC},k} \le Q_{\max}. 
    \label{eq:hvac-capacity}
\end{align}
To avoid rapid fluctuations in the HVAC operation, we impose ramp-rate constraints,
\begin{align}
    Q_{\text{HVAC},k} - Q_{\text{HVAC},k-1} &\le R_{\uparrow}, \\
    Q_{\text{HVAC},k-1} - Q_{\text{HVAC},k} &\le R_{\downarrow}, \qquad \forall\, k \ge 1,
    \label{eq:hvac-ramp}
\end{align}

Together with the discrete-time dynamics~\eqref{eq:building-ss}, these constraints define the feasible set of building trajectories over the MPC horizon. A sampling period of $\tau = 1$\,h is adopted throughout, consistent with the slow thermal time constants of residential buildings (typically several hours), the resolution of day-ahead electricity markets, and the hourly scheduling of battery and grid decisions. Sub-hourly thermal transients are negligible at the MPC level, justifying the joint optimisation of thermal and power dynamics at this timescale.

The internal heat gains $Q_{\text{ihg}}$ are considered known and constant, although they can be extended to include stochastic variations. The effective solar gains $q_{\text{sol}}A$ are computed using the solar geometry approach of (\cite{Buenning2022}), which accounts for both solar elevation and the relative orientation of the window to the sun. For each time step $k$, the effective solar heat gain $Q_{\text{sol},k}$ [kW] is given by
\begin{align}
Q_{\text{sol},k} 
    &= I_{\text{hor},k}\, A_{\text{win}}\,
       \sin^+\!(\theta_{\text{el},k})\,
       \cos^+\!\big(\theta_{\text{az},k} - \alpha_{\text{win}}\big), \label{eq:solar-gains-bunning}
\end{align}
where $I_{\text{hor},k}$ denotes the global horizontal irradiance, $\theta_{\text{el},k}$ and $\theta_{\text{az},k}$ are the solar elevation and azimuth angles, $\alpha_{\text{win}}$ the window azimuth, $A_{\text{win}}$ the effective window area, and $\sin^+(\cdot) = \max(0,\sin(\cdot))$, $\cos^+(\cdot) = \max(0,\cos(\cdot))$. Irradiance is not known in advance and is forecasted over the MPC horizon. Since the electricity price enters the global objective via the power balance, we use a purely comfort-driven building cost to avoid double counting:
\begin{align}
    J_{\text{house}}
        = \sum_{k=0}^{H-1}
         \bigl(T_{z,k} - T_{z,k}^{\text{set}}\bigr)^{2},
    \label{eq:building-cost}
\end{align}
where $T_{z,k}^{\text{set}}$ is the indoor temperature setpoint. Although the building model does not explicitly optimize electricity cost, the HVAC actuation directly affects the VPP-wide grid exchange, implicitly incentivizing demand shifting when prices are high.

\subsection{Electricity Grid Connection}
The VPP interacts with the electricity grid through electricity prices denoted by~\(\lambda_k\). We assume that the electricity is traded at the day-ahead price which we treat as an uncertain signal in this work. To account for transaction costs and market spread, the effective buy and sell prices at time~\(k\) are defined as
\begin{align}
\lambda^{\text{buy}}_k &= (1+\delta_{\text{s}})\,\lambda_k, \qquad
\lambda^{\text{sell}}_k = \beta_{\text{sell}}\,\lambda_k,
\label{eq:vpp-price}
\end{align}
with \(\delta_{\text{s}}>0\) denoting the relative price markup for energy imports and \(\beta_{\text{sell}}\in(0,1)\) the relative discount for energy exports. 
This ensures that \(\lambda^{\text{sell}}_k < \lambda_k < \lambda^{\text{buy}}_k\).

\subsection{Optimization Problem for Cost Minimization and User Comfort}
The VPP solves at each time step $\kappa$ a finite-horizon optimization problem jointly minimizing energy cost, thermal discomfort, and battery degradation:
\begin{multline}
\min_{\{u_k,x_k\}_{k=\kappa}^{\kappa+H}} \;
J
=
\sum_{k=\kappa}^{\kappa+H-1}
\tau\Bigl(
\lambda^{\mathrm{buy}}_k\, P_{\mathrm{imp},k}
-
\lambda^{\mathrm{sell}}_k\, P_{\mathrm{exp},k}
\Bigr)
\\[1ex]
\quad
+ \;
w_T
\sum_{k=\kappa}^{\kappa+H-1}
\bigl(T_{z,k}-T^{\mathrm{set}}_{z,k}\bigr)^{2}
+
\sum_{k=\kappa}^{\kappa+H-1}
c_{\deg}\,\bigl(P^{\mathrm{ch}}_k + P^{\mathrm{dis}}_k\bigr)\,\tau
.
\label{eq:vpp-cost-weighted}
\end{multline}
The first term captures market revenue, the second penalizes comfort deviations, and the third introduces a degradation cost proportional to battery throughput. The weight $w_T$ scales the comfort penalty relative to trading profit and degradation cost.

\section{Forecasting and Uncertainty}
\label{uncertainty}

\begin{figure}[ht!]
    \centering
    \includegraphics[width=1\linewidth]{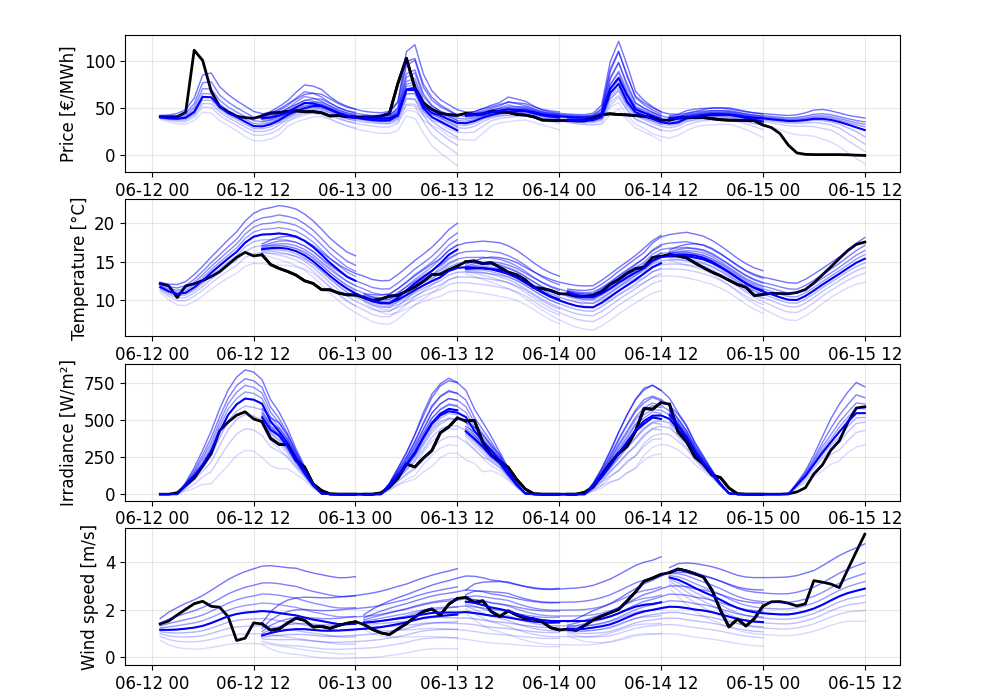}
    \caption{Probabilistic forecasts of all uncertain quantities. Blue lines show 
    forecasts issued at 12-hour intervals; light blue lines show the corresponding 
    quantiles.}
    \label{fig:placeholder}
\end{figure}

Accurate short-term probabilistic forecasts of renewable generation, demand, and 
electricity prices are essential for uncertainty-aware VPP operation 
(\cite{Morales2014, Gneiting2014}). In this work, we use the Time-series Representation Extractor (TiREx) forecaster 
(\cite{Auer2025}) to generate probabilistic predictions for electricity price, 
ambient temperature, solar irradiance, and wind speed in a purely causal fashion. 
TiREx is a pre-trained zero-shot model based on an xLSTM architecture that 
generalizes to unseen time series through large-scale pre-training, allowing 
immediate inference in new domains. Given a configurable lookback window $y_{1:T}$ (we use $T=10$ days), it outputs conditional quantiles $\{\hat{y}^q_t : q \in Q\}$ with $Q = \{0.1, 0.2, \ldots, 0.9\}$ over a 24-hour horizon, as illustrated in Figure~\ref{fig:placeholder}.  For solar irradiance, TiREx forecasts the clearness index $k_t = I_{\text{sol},t} / I_{\text{cs},t}$ and back-transforms via the deterministic clear-sky irradiance at the Oslo location.

While perfect forecasts would yield optimal decisions, they are unattainable in 
practice due to the stochastic nature of weather, demand, and electricity prices 
(\cite{Morales2014}). A common method to incorporate uncertainty is scenario-based stochastic programming (SP), representing uncertain variables through sampled realizations (\cite{Campi2009}). Although SP offers probabilistic guarantees, capturing temporal and cross-variable dependencies in VPP applications typically requires a large number of scenarios, substantially increasing computational burden. Motivated by these challenges, we instead construct stage-wise empirical predictive distributions using quantile forecasts, providing a lightweight and data-driven representation of uncertainty. 

\begin{figure}[ht!]
    \centering
    \includegraphics[width=0.9\linewidth]{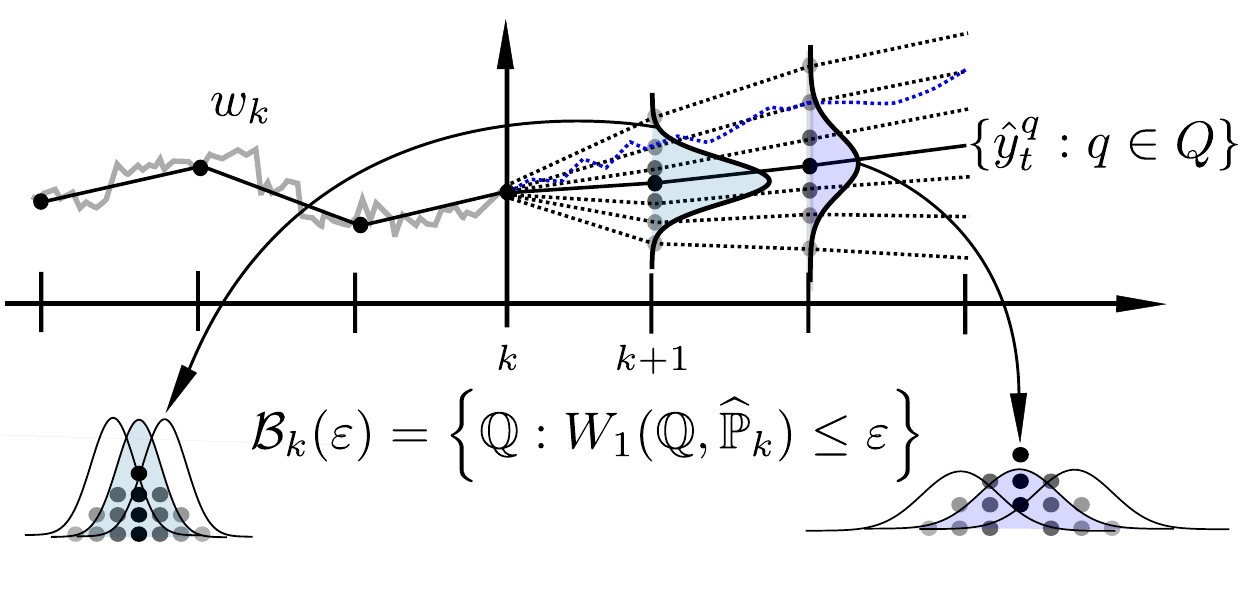}
    \caption{From the forecasted quantiles, data-driven Wasserstein ambiguity sets 
    are constructed and incorporated into the decision-making process.}
    \label{fig:algorithm}
\end{figure}

At each time step $k$, the forecasting model provides predictive quantiles 
$\bigl\{\widehat{y}^{\,q}_k : q \in Q \bigr\}$ with $Q = \{q_1,\ldots,q_m\}$, 
$q_i \in (0,1)$, as illustrated in Figure~\ref{fig:algorithm}. 

Each quantile is assigned a probability weight $w(q) \geq 0$ with $\sum_{q\in Q} w(q) = 1$; triangular weights can be used to concentrate mass around the central quantiles while uniform quantiles assign equal probability to each of the quantile values. The stage-wise empirical distribution at time $k$ is then defined as
\[
    \widehat{\mathbb{P}}_k
    :=
    \sum_{q \in Q} 
        w(q)\,\delta_{\widehat{y}^{\,q}_k},
\]
where $\delta_x$ denotes the Dirac measure at $x$. We construct these distributions independently for each prediction step $k$, 
since the quantile forecasts provide marginal rather than jointly calibrated 
information across the horizon. Importantly, this stage-wise 
reconstruction is what renders the ambiguity sets time-varying: the center 
$\widehat{\mathbb{P}}_k$ of each Wasserstein ball adapts to the current forecast 
dispersion at each stage, while the radius $\varepsilon$ remains fixed throughout 
the horizon.

\section{Distributionally Robust Model Predictive Control}
\label{control}

The overall VPP model can be compactly written as a discrete-time dynamical system
\begin{align}
x_{k+1} = f(x_k, u_k, w_k), \label{eq:sysdyn}
\end{align}
where $x_k$ denotes the system state (indoor temperatures and battery SoC), 
$u_k$ the control input (HVAC power, charging/discharging, grid exchange), 
and $w_k$ uncertain exogenous variables including temperature, irradiance, wind 
speed, and electricity prices. Since key components of the VPP objective---most 
notably electricity prices---are uncertain, we reformulate the finite-horizon 
optimization as a DR-MPC problem using time-varying Wasserstein ambiguity sets 
constructed from the TiREx probabilistic forecasts.

\subsection{Nominal MPC formulation}
For the MPC formulation it is convenient to define the stage cost as
\begin{equation}
\begin{aligned}
J(x_k,u_k,w_k) 
:=& -\big(\lambda^{\mathrm{sell}}_k P_{\mathrm{exp},k}
       - \lambda^{\mathrm{buy}}_k P_{\mathrm{imp},k}\big)\,\tau \\ 
   &+ w_T\big(T_{z,k}-T^{\mathrm{set}}_{z,k}\big)^{2} 
   + c_{\mathrm{deg}}\big(P^{\mathrm{ch}}_{k}+P^{\mathrm{dis}}_{k}\big)\,\tau
\end{aligned}
\label{eq:stage-cost}
\end{equation}
such that minimizing the cumulative cost is equivalent to maximizing the profit
in~\eqref{eq:vpp-cost-weighted}. In a nominal MPC framework, the control sequence 
$u_{\kappa:\kappa+H-1}$ is obtained by minimizing
\begin{align}
\min_{u_{\kappa:\kappa+H-1}} \; 
\sum_{k=\kappa}^{\kappa+H-1} J(x_k,u_k,\hat w_k),
\label{eq:nominal-mpc}
\end{align}
subject to system dynamics, power balance, and operational constraints on 
temperatures, battery SoC, and power limits, where $\hat w_k$ denotes the 
point forecast of the disturbance at time~$k$.

\subsection{Stage cost as an affine function of the price}
The price-dependent part of the stage cost is affine in $\lambda_k$, allowing us 
to write $J_k(x_k,u_k,\lambda_k) = a_k(x_k,u_k)\,\lambda_k + b_k(x_k,u_k)$, 
where
\[
a_k(x_k,u_k)
=
\tau\Big((1+\delta_{\mathrm{s}})P_{\mathrm{imp},k}
-
\beta_{\mathrm{sell}}P_{\mathrm{exp},k}\Big),
\]
and $b_k$ collects all price-independent terms. The distributionally robust formulation is applied exclusively to electricity  price uncertainty. As shown above, the stage cost $J_k$ is affine in the market price $\lambda_k$, which enables the tractable closed-form reformulation of the worst-case expectation. Incorporating uncertainty in temperature or solar irradiance would introduce nonlinear dependencies through the building thermal dynamics and the PV generation model, breaking the affine structure and precluding a similarly tractable reformulation.

\subsection{Wasserstein Ambiguity set and DRO reformulation}

For each stage~$k$, we construct an $\ell_1$--Wasserstein ball
\[
\mathcal{B}_k(\varepsilon)
=
\Bigl\{
\mathbb{Q} : W_1(\mathbb{Q},\widehat{\mathbb{P}}_k)\le \varepsilon
\Bigr\},
\]
centered at the empirical price distribution $\widehat{\mathbb{P}}_k$.
Since $J_k$ is affine in the uncertain price~$\lambda_k$, the worst-case 
expected stage cost admits the closed-form representation (\cite{MohajerinEsfahani2017})
\[
\sup_{\mathbb{Q}_k\in\mathcal{B}_k(\varepsilon)}
\mathbb{E}_{\mathbb{Q}_k}[J_k]
=
\mathbb{E}_{\widehat{\mathbb{P}}_k}[J_k]
+
\varepsilon\,|a_k(x_k,u_k)|.
\]
The DRO correction term penalizes the price sensitivity
\(
a_k(x_k,u_k)
\),
which is proportional to the net exposure to the electricity market. The radius $\varepsilon$ controls the conservatism of the distributionally robust formulation: larger values increase robustness against distributional shift at the cost of reduced expected profit. In practice, $\varepsilon$ can be selected via cross-validation or backtesting on historical data, by evaluating the trade-off between robustness and economic performance across a candidate set of radii. Data-driven methods that provide finite-sample coverage guarantees offer an alternative principled selection procedure (\cite{MohajerinEsfahani2017}). In this work, $\varepsilon \in \{0.25, 0.5, 1, 2\}$ is evaluated empirically, as discussed in Section~\ref{case}.

\section{Case Study and Results}
\label{case}
To demonstrate the applicability of the proposed control framework, we consider a simulated VPP representative of a small neighborhood in Oslo, Norway, aggregating $N_h$ identical residential buildings with HVAC systems, $N_b$ identical shared batteries, and local solar and wind generation, with full state measurement access.

\subsection{Data and Experimental Setup}

Weather data (temperature, irradiance, wind speed) are obtained from the ERA5 reanalysis dataset (\cite{ERA5}), and day-ahead electricity prices from the ENTSO-E Transparency Platform for bidding zone NO1 (\cite{ENTSOE}), which are treated as exogenous signals determining the economic incentives for grid import and export. Simulations are performed over two continuous 30-day periods in 2024: a spring period (15~Apr--15~May) with moderate temperatures and significant solar generation, and an autumn period (1--31~Oct) with increased heating demand and reduced irradiance. At each step, the controller solves a 24-hour MPC problem, applies the first input, and advances the horizon by one hour. All optimization problems were solved using MOSEK as the primary solver, with OSQP and ECOS as fallback solvers, via the CVXPY modelling framework. Probabilistic forecasts were precomputed offline, and the resulting quadratic programs were solved within seconds at each MPC step, remaining well within the one-hour sampling interval.

We compare three control strategies:
\begin{itemize}
    \item \textbf{PF-MPC:} assumes perfect knowledge of future signals, 
    serving as an upper bound on achievable revenue.
    \item \textbf{FC-MPC:} uses TiREx point forecasts without robustness 
    margins, representing a realistic baseline.
    \item \textbf{DR-MPC:} augments FC-MPC with a distributionally robust 
    price term using a Wasserstein ambiguity set of radius~$\varepsilon$.
\end{itemize}

For each period we consider two uncertainty regimes: \emph{full-uncertainty}, where all stochastic inputs are forecasted, and \emph{price-only}, where weather is known perfectly and only price uncertainty remains. This separation isolates the financial impact of price uncertainty from physical uncertainty in building and renewable dynamics. To construct the empirical price distributions, this work uses triangular weights. Detailed simulation parameters are provided in Table~\ref{tab:parameters} in the Appendix.

\subsection{Evaluation}
\begin{figure}[ht!]
    \centering
    \includegraphics[width=0.95\linewidth]{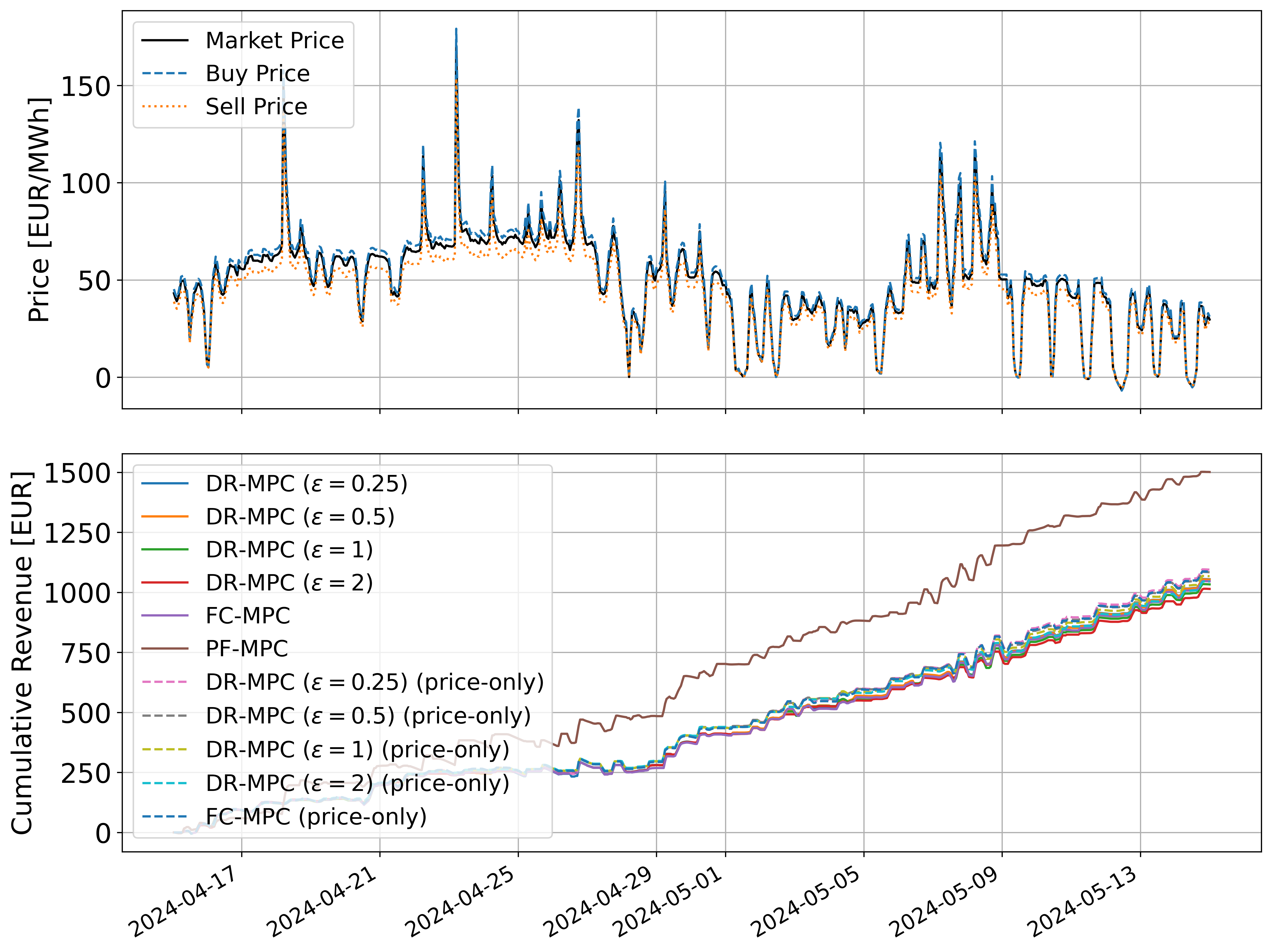}
    \caption{Prices and cumulative revenue in spring scenario.}
    \label{fig:spring_scenario}
\end{figure}

\begin{figure}[ht!]
    \centering
    \includegraphics[width=0.95\linewidth]{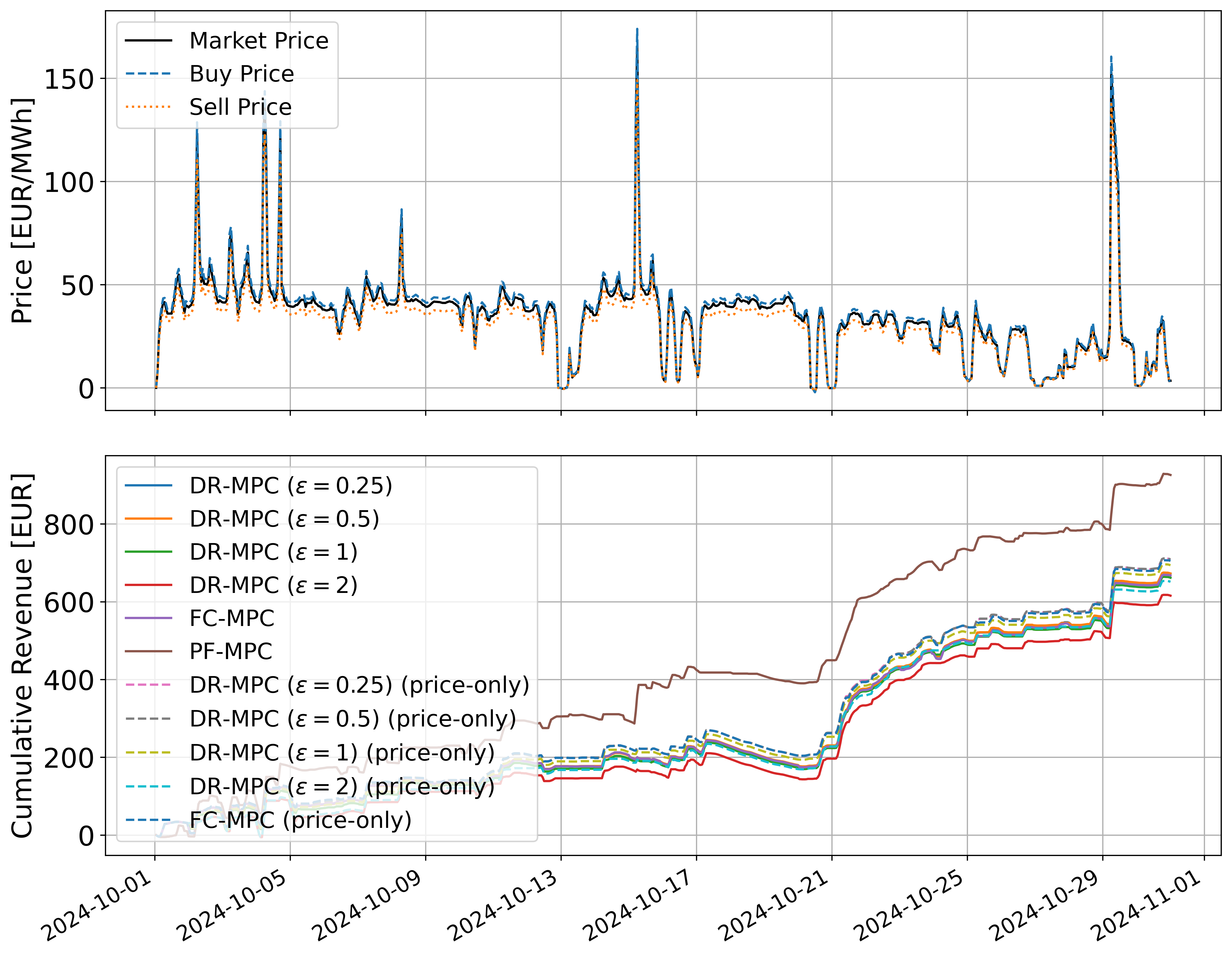}
    \caption{Prices and cumulative revenue in autumn scenario.}
    \label{fig:autumn_scenario}
\end{figure}

As performance metrics we report the cumulative revenue and relative improvement over FC-MPC across both 30-day periods for $\varepsilon \in \{0.25, 0.5, 1, 2\}$. Across all experiments, PF-MPC achieves the highest cumulative revenue, as expected from its access to perfect future information. It therefore serves solely as an upper bound and is not directly comparable to the forecast-based strategies. The results of both evaluation periods are summarized in Figures \ref{fig:spring_scenario}–\ref{fig:autumn_scenario} and Tables \ref{tab:rev_spring}–\ref{tab:rev_autumn}. 

In the spring scenario (15~Apr--15~May~2024), only small ambiguity radii improve performance over FC-MPC, with the best-performing variant ($\varepsilon = 0.25$) achieving a gain of +0.82\,\% under full uncertainty and +0.78\,\% in the price-only case. Larger radii become increasingly conservative and reduce revenue.

\begin{table}[ht!]
\caption{Cumulative revenue for the spring period (15~Apr--15~May~2024) and
relative improvement over FC-MPC.}
\label{tab:rev_spring}
\centering
\scriptsize
\renewcommand{\arraystretch}{1.15}

\begin{tabular}{c|cc|cc}
\hline
& \multicolumn{2}{c|}{\textit{full-uncertainty}} 
& \multicolumn{2}{c}{\textit{price-only}} \\
& Rev [EUR] & $\Delta$ [\%] 
& Rev [EUR] & $\Delta$ [\%] \\
\hline
PF-MPC                         & 1502 & --      & 1502 & --      \\
FC-MPC                         & 1046 & 0.0     & 1086 & 0.0     \\
DR-MPC ($\varepsilon{=}0.25$)  & 1055 & +0.82   & 1095 & +0.78   \\
DR-MPC ($\varepsilon{=}0.5$)   & 1053 & +0.68   & 1084 & $-0.22$ \\
DR-MPC ($\varepsilon{=}1$)     & 1034 & $-1.21$ & 1068 & $-1.74$ \\
DR-MPC ($\varepsilon{=}2$)     & 1015 & $-3.02$ & 1045 & $-3.81$ \\
\hline
\end{tabular}
\end{table}
The autumn scenario is characterized by lower solar generation, colder temperatures, and reduced price volatility, all of which limit arbitrage opportunities and increase heating demand. As a result, overall revenue levels are lower compared to the spring experiment. 

The autumn scenario (1--31~Oct~2024) shows a consistent pattern with spring: small ambiguity radii improve performance, with $\varepsilon = 0.5$ performing 
best at +0.76\,\% under full uncertainty and +0.57\,\% in the price-only case, while larger radii lead to pronounced performance drops reaching $-7.77\,\%$ for 
$\varepsilon = 2$.

\begin{table}[ht!]
\caption{Cumulative revenue for the autumn period (1--31~Oct~2024) and
relative improvement over FC-MPC.}
\label{tab:rev_autumn}
\centering
\scriptsize
\renewcommand{\arraystretch}{1.15}
\begin{tabular}{c|cc|cc}
\hline
& \multicolumn{2}{c|}{\textit{full-uncertainty}} 
& \multicolumn{2}{c}{\textit{price-only}} \\
& Rev [EUR] & $\Delta$ [\%] 
& Rev [EUR] & $\Delta$ [\%] \\
\hline
PF-MPC                         & 927  & --      & 927  & --      \\
FC-MPC                         &  669 & 0.0     &  706 & 0.0     \\
DR-MPC ($\varepsilon{=}0.25$)  &  673 & +0.69   &  710 & +0.51   \\
DR-MPC ($\varepsilon{=}0.5$)   &  674 & +0.76   &  710 & +0.57   \\
DR-MPC ($\varepsilon{=}1$)     &  663 & $-0.88$ &  695 & $-1.59$ \\
DR-MPC ($\varepsilon{=}2$)     &  617 & $-7.77$ &  652 & $-7.61$ \\
\hline
\end{tabular}
\end{table}

Larger ambiguity radii ($\varepsilon = 2$) in comparison to moderate selections ($\varepsilon = 1$) lead to a pronounced performance drop, consistent with the intuition that overly conservative distributions restrict profitable arbitrage, while removing weather uncertainty increases absolute revenue levels across both scenarios.

DR-MPC improves revenue when the ambiguity radius is chosen appropriately, with consistent gains of up to 0.8\,\% across both scenarios for small radii. Both spring and autumn show the same qualitative pattern: small radii provide modest 
but consistent improvements, while larger radii become overly conservative and restrict profitable arbitrage. The price-only experiments confirm that price uncertainty is the primary driver of economic risk, while removing weather uncertainty mainly increases absolute revenue levels. Overall, DR-MPC provides a simple and effective enhancement of standard MPC, offering consistent economic benefits with only a minor adjustment to the objective.

\section{Conclusion}
\label{conclusion}
This paper presented a DR-MPC framework for real-time VPP operation under forecast uncertainty, combining quantile-based Wasserstein ambiguity sets with a tractable reformulation that seamlessly extends standard MPC. A case study using real Norwegian data demonstrated potential revenue gains of up to 0.8\,\% over forecast-based MPC for small radii.

A limitation of the proposed framework concerns the stage-wise construction of the empirical price distributions 
$\widehat{\mathbb{P}}_k$. Although the underlying quantile forecasts are generated jointly by the same model conditioned 
on a common lookback window, the resulting ambiguity sets $\mathcal{B}_k(\varepsilon)$ are constructed independently at 
each prediction step from marginal quantile forecasts, and temporal dependencies in the price process are therefore not 
explicitly captured. In the multi-step MPC setting with battery storage, where optimal charge and discharge decisions depend on 
the joint distribution of prices across the horizon, this stage-wise independence may lead to overly conservative 
robustification of intertemporal arbitrage decisions. Constructing joint Wasserstein ambiguity sets over price 
trajectories, for instance via scenario-based representations of temporal dependencies represents 
an important direction for future work.

Future research will also focus on extending the VPP model to heterogeneous building and battery portfolios, extending the distributionally robust framework to incorporate additional uncertain signals, and developing data-driven or scheduled mechanisms for selecting the Wasserstein radius according to the season or time. Furthermore, uncertainty in renewable generation and thermal demand could be incorporated through sample average approximation or scenario-based stochastic programming, complementing the distributionally robust treatment of price uncertainty within a unified framework. Strict mutual exclusivity of import and export could additionally be enforced via mixed-integer programming.

\noindent\textbf{Declaration of Generative AI and AI Technologies in the Writing Process:}
During the preparation of this work, the author(s) used ChatGPT in order to
improve the grammar of this paper. After using this tool/service, the author(s)
reviewed and edited the content as needed and take full responsibility for the
content of the publication.

\bibliography{PhD_Sources}             

\newpage
\section{Appendix}
\begin{table}[ht]
\centering
\caption{Model Parameters and Symbols}
\label{tab:parameters}
\resizebox{\columnwidth}{!}{%
\begin{tabular}{llcc}
\toprule
\textbf{Symbol} & \textbf{Description} & \textbf{Value} & \textbf{Unit} \\
\midrule
\multicolumn{4}{l}{\textit{Building Model (3R2C)}} \\[2pt]
$C_1,\,C_2,\,C_z$ & Thermal capacitances (outer, inner wall, zone) & 25,\;80,\;3 & kWh/°C \\
$R$ & Wall thermal resistance & 2 & °C/kW \\
$hA$ & Heat transfer coefficient & 2 & kW/°C \\
$Q_{\text{ihg}}$ & Internal heat gains (daily profile) & 0.3--0.8 & kW \\
$A_{\text{win}}$ & Effective window area & 15 & m$^2$ \\
$\alpha_{\text{win}}$ & Window azimuth angle & 180 & ° \\
$T_{z,k}^{\text{set}}$ & Indoor temperature setpoint & 20 & °C \\
$T_{z}^{\min},\,T_{z}^{\max}$ & Zone temperature limits & 15--25 & °C \\
$w_T$ & Thermal comfort weighting factor & 100 & -- \\
COP & Coefficient of performance (HVAC) & 3 & -- \\
$Q_{\max}$ & Maximum HVAC thermal power & 10 & kW \\
$R_{\text{up}},\,R_{\text{down}}$ & HVAC ramp-rate limits & 2 & kW/h \\
\\[-3pt]
\multicolumn{4}{l}{\textit{Battery Model}} \\[2pt]
$\eta^{\text{ch}},\,\eta^{\text{dis}}$ & Charging/discharging efficiencies & 0.95 & -- \\
$P^{\text{ch}}_k,\,P^{\text{dis}}_k$ & Charging/discharging powers & 0--20 & kW \\
$C_{\mathrm{bat}}$ & Battery capacity & 100 & kWh \\
$S^{\min},\,S^{\max}$ & SoC limits (fraction of capacity) & 0.2,\;0.9 & -- \\
$\bar P^{\text{ch}},\,\bar P^{\text{dis}}$ & Power limits & 20 & kW \\
$c_{\mathrm{deg}}$ & Degradation cost coefficient & 0.01 & -- \\
\\[-3pt]
\multicolumn{4}{l}{\textit{Renewable Generation}} \\[2pt]
$A_{\text{PV}}$ & Total PV area & 200 & m$^2$ \\
$\eta_{\text{PV}}$ & PV module efficiency & 0.18 & -- \\
$\gamma_T$ & PV temperature coefficient & 0.0045 & 1/°C \\
$k_T$ & PV temperature correction coefficient & 0.03 & °C·m$^2$/W \\
$\rho_{\text{air}}$ & Air density & 1.225 & kg/m$^3$ \\
$r_{\text{b}}$ & Wind turbine rotor radius & 15 & m \\
$\eta_{\text{wind}}$ & Wind conversion efficiency & 0.45 & -- \\
\\[-3pt]
\multicolumn{4}{l}{\textit{Grid Connection}} \\[2pt]
$\Delta_{\text{spread}}$ & Buy--sell price spread & 0.05 & €/kWh \\
$\beta_{\text{sell}}$ & Relative selling price factor & 0.9 & -- \\
\\[-3pt]
\multicolumn{4}{l}{\textit{Simulation Parameters}} \\[2pt]
$T$ & Lookback horizon TiREx & 10 & days \\
$N_{\text{h}}$ & Number of buildings & 25 & -- \\
$N_{\text{bat}}$ & Number of batteries & 15 & -- \\
$N_{s}$ & Number of PV units & 5 & -- \\
$N_{w}$ & Number of wind turbines & 7 & -- \\
\bottomrule
\end{tabular}
}%
\end{table}

\end{document}